\documentclass[conference]{IEEEtran}
\IEEEoverridecommandlockouts
\usepackage{cite}
\usepackage{amsmath,amssymb,amsfonts}
\usepackage{algorithmic}
\usepackage{graphicx}
\usepackage{textcomp}
\usepackage{xcolor}
\usepackage{breakurl}
\usepackage{mathtools}
\usepackage[caption=false]{subfig}
\usepackage{flushend}

\usepackage{soul}

\makeatletter
\newcommand{\linebreakand}{%
  \end{@IEEEauthorhalign}
  \hfill\mbox{}\par
  \mbox{}\hfill\begin{@IEEEauthorhalign}
}
\makeatother

\def\BibTeX{{\rm B\kern-.05em{\sc i\kern-.025em b}\kern-.08em
    T\kern-.1667em\lower.7ex\hbox{E}\kern-.125emX}}
\begin{document}

\title{Deep Unfolded Approximate Message Passing for Quantitative Acoustic Microscopy Image Reconstruction\\
\thanks{This work was supported by NIH grant R01GM143388.}
}

\author{\IEEEauthorblockN{Odysseas Pappas}
\IEEEauthorblockA{\textit{Visual Information Laboratory} \\
\textit{University of Bristol}\\
Bristol, UK \\
o.pappas@bristol.ac.uk}
\and
\IEEEauthorblockN{Jonathan Mamou}
\IEEEauthorblockA{\textit{Department of Radiology} \\
\textit{Weill Cornell Medicine}\\
New York, USA \\
jom4032@med.cornell.edu}
\and
\IEEEauthorblockN{Adrian Basarab}
\IEEEauthorblockA{\textit{CREATIS} \\
\textit{University of Lyon}\\
Lyon, France \\
adrian.basarab@creatis.insa-lyon.fr}
\linebreakand 
\IEEEauthorblockN{Denis Kouam\'{e}}
\IEEEauthorblockA{\textit{IRIT} \\
\textit{University of Toulouse III, Paul Sabatier }\\
Toulouse, France \\
denis.kouame@irit.fr}
\and
\IEEEauthorblockN{Alin Achim}
\IEEEauthorblockA{\textit{Visual Information Laboratory} \\
\textit{University of Bristol}\\
Bristol, UK \\
alin.achim@bristol.ac.uk}
}

\maketitle

\begin{abstract}
Quantitative Acoustic Microscopy (QAM) is an imaging technology utilising high frequency ultrasound to produce quantitative two-dimensional (2D) maps of acoustical and mechanical properties of biological tissue at microscopy scale. Increased frequency QAM allows for finer resolution at the expense of increased acquisition times and data storage cost. Compressive sampling (CS) methods have been employed to produce QAM images from a reduced sample set, with recent state of the art utilising Approximate Message Passing (AMP) methods. In this paper we investigate the use of AMP-Net, a deep unfolded model for AMP, for the CS reconstruction of QAM parametric maps. Results indicate that AMP-Net can offer superior reconstruction performance even in its stock configuration trained on natural imagery (up to 63$\%$ in terms of PSNR), while avoiding the emergence of sampling pattern related artefacts.
\end{abstract}


\section{Introduction}
Quantitative Acoustic Microscopy (QAM) is a relatively new imaging technology for the investigation of soft biological tissue, capable of producing quantitative maps describing the acoustical and mechanical properties of tissue at microscopic resolution \cite{Mamou2017} \cite{Mamou2017B}. QAM systems utilise tightly focused beams of high frequency ultrasound (250 MHz to 1 GHz), transmitted in a 2-D raster-scanning fashion over the tissue sample of interest. By processing the received radio-frequency (RF) signals, it is possible to obtain parametric maps of a number of the tissue's acoustical and mechanical properties such as speed of sound, acoustic impedance, bulk modulus and mass density. This novel contrast mechanism provides complementary information to typical histology photomicrographs and optical or electron microscopy and can be of great value in a clinical and diagnostic context \cite{Hildebrand1981}.  


Increased resolution for QAM systems requires the use of higher-frequency transducers and comes at the cost of increased acquisition times and increased data storage and processing costs, as the raster scanning grid over the sample becomes finer in resolution. This has motivated work along a number of avenues including super-resolution methods for QAM \cite{Dutta2024}, as well as possible ways to reconstruct parametric maps from a spatially reduced set of measurements \cite{Kim2017} or sparsely sampled RF echoes \cite{Kim2020}.

In this context, compressed sensing (CS) methods have become of particular interest to QAM. CS methods are inherently tied to the concept of sparsity and form a mathematical framework under which a signal can be successfully reconstructed from a parsimonious set of measurements, typically numbering far fewer total samples than what the Nyquist theorem would define as the required minimum for reconstruction \cite{Donoho2006} \cite{Candes2008}. Motivated by previous work utilising Approximate Message Passing (AMP) for the CS reconstruction of QAM maps \cite{Kim2017} \cite{Kim2018}, we investigate here the use of AMP-Net, a network mimicking AMP via the process of deep unfolding. 


\section{Background}
\subsection{Quantitative Acoustic Microscopy}
\begin{figure}[t]
\centering
\includegraphics[width=.35\textwidth]{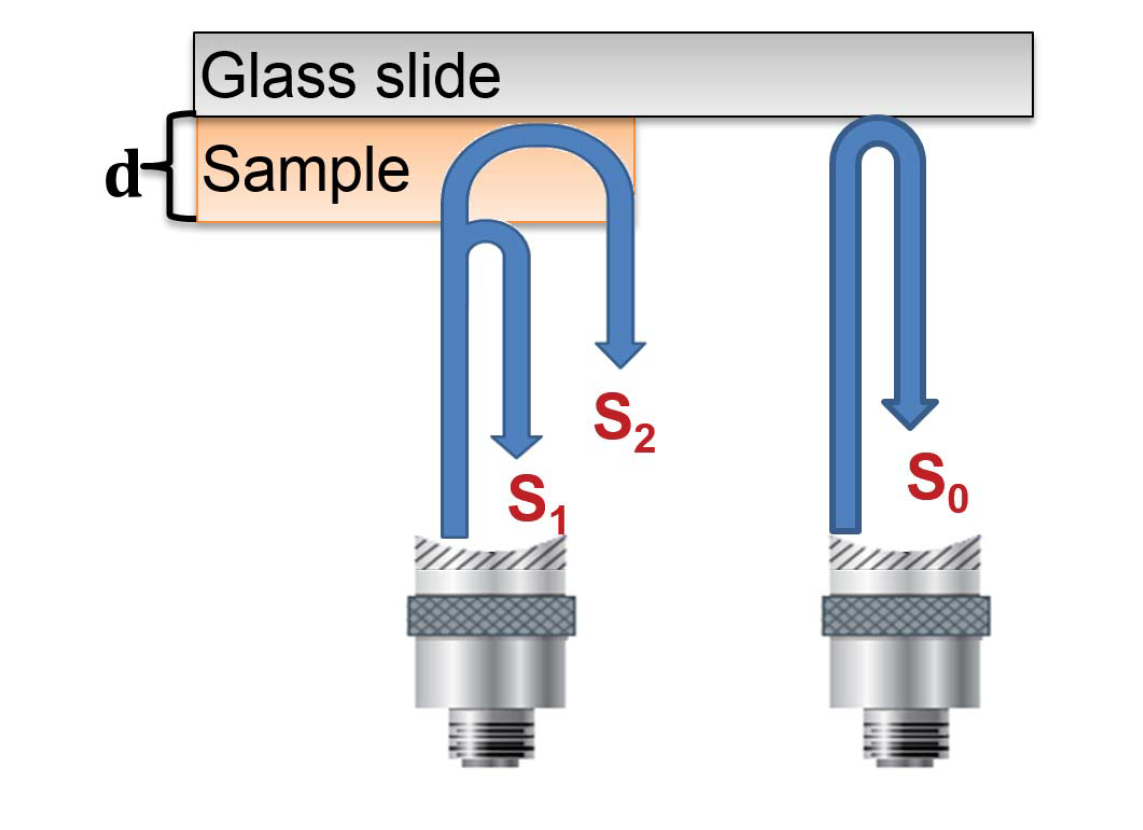}
\caption{Working principle of QAM. RF data from areas of the glass slide containing the sample being imaged consist of two primary reflections, one occurring at the coupling medium-sample interface ($S_1$) and one at the sample-glass interface ($S_2$). Over parts of the glass slide that contain no sample, the RF signal contains only one reflection ($S_0$)}
\label{fig:reflections}
\end{figure}

QAM relies on the emission of a tightly focused, high-frequency ($>$250 MHz) ultrasound pulses over part of the sample of interest, typically a thin section of soft tissue affixed to a glass slide. This process is repeated in a 2D raster scanning fashion so as to cover the entirety of the sample of interest. The recorded RF echoes consist primarily of two reflections; one ($S_1$) from the interface between the coupling medium and the sample, and one ($S_2$) from the interface between the sample and the glass slide. Over parts of the glass slide that do not contain a sample, we obtain a single-reflection reference signal $S_0$ (see Figure \ref{fig:reflections}).

\begin{equation}
S(t) = S_1(t) + S_2(t)
\label{reflectioneq}
\end{equation}

The echo signals $S_1(t)$ and $S_2(t)$ can be expressed as amplitude decayed and time delayed versions of the reference signal $S_0(t)$ as 

\begin{equation}
S(t) = a_1S_0(t-t_1) + a_2S^*_0(t-t_2)
\label{reflectioneq2}
\end{equation}

where $a_{1,2}$ are amplitude decays and $t_{1,2}$ are time delays, with $(*)$ denoting an additional effect of frequency dependent attenuation. Knowledge of these parameters allows for the computation of the tissue's acoustic and mechanical properties including the speed of sound (SoS), acoustic impedance, and ultrasound attenuation coefficient \cite{Rohrbach2018}. 

In a QAM system the entirety of the sample slide must be scanned in a 2D raster fashion, acquiring one RF signal per position in the sample and forming a complete RF data cube for the entire sample. This process of physically moving the transducer along the grid positions is one of the major bottlenecks in QAM data acquisition. Being able to reconstruct parameter maps from a subsampled set of spatial measurements brings obvious advantages in terms of both acquisition time and data storage requirements, both of which can become major concerns with ultra-high resolution 1 GHz systems. 

\subsection{Approximate Message Passing CS Reconstruction for QAM}
CS theory \cite{Donoho2006} posits that if a signal $ x \in \mathbb{R}^N $ is $K$-sparse in one basis, .ie. is representable by $K$ elements in this basis, then it can be recovered or reconstructed from a total of $M = cst \cdot K \cdot log(N/K) << N$ linear projections onto a second measurement basis, where $cst$ is a small overmeasuring constant typically set $>1$. 

The measurement model therefore becomes $y = Ax + n$ where $y \in \mathbb{R}^M$ is the measurement vector, $x$ is the signal to be reconstructed, $A \in \mathbb{R}^{M \times N}$ is the measurement matrix and $n$ is an additive noise term. Reconstruction of $x$ is possible via a number of approaches often involving constrained optimisation problems, such as Orthogonal Matching Pursuit \cite{Tropp2007}, or iterative thresholding approaches such as Approximate Message Passing (AMP) \cite{Donoho2010} \cite{Tan2015}. 


AMP image reconstruction can be thought of as an iterative denoising process, where successive removals of noise lead to an image with acceptable noise variance. The process can be summarized as follows: 

\begin{equation}
x^{k} = \mathfrak{T}_k(A^Tz^{k-1} + x^{k-1})
\label{amp1}
\end{equation}
\begin{equation}
z^{k-1} = y - Ax^{k-1}
\label{amp2}
\end{equation}

where $z$ denotes the residual signal, $A$ is the measurement matrix and $\mathfrak{T}(\cdot)$ is the denoiser function. Superscripts $k$ and $T$ correspond to the iteration number and transpose respectively. 

\begin{figure*}[t]
\centering
\includegraphics[width=0.75\textwidth]{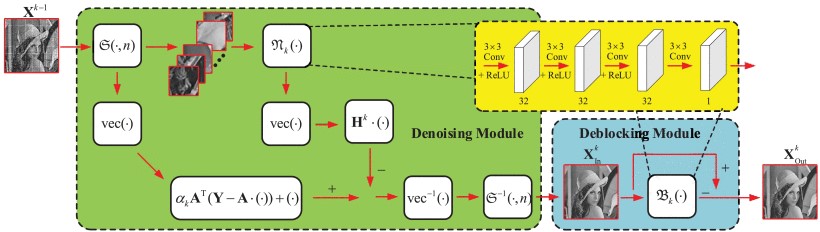}
\caption{Structure of the $k$-th iteration of the reconstruction module of AMP-Net. Reproduced from \cite{Zhang2021}.}
\label{fig:ampnetstruct}
\end{figure*}

AMP has been employed in the past to reconstruct sparsely sampled QAM images. In \cite{Kim2018}, the authors present an extended wavelet-based AMP approach for QAM imaging that utilises a Cauchy \textit{maximum a posteriori} image denoising algorithm that accounts for the non-Gaussianity of QAM wavelet coefficients \cite{Hill2016}. In \cite{Kim2018} the authors also proposed a series of binary spatial sampling patterns, with the most promising performance shown by a spiral pattern. While this makes for an practically realisable pattern, it is theoretically flawed as it does not allow for the measurement of linear combinations of samples but is rather a binary matrix, denoting whether data acquisition over a particular point of the sample is to be made or not. This spiral sampling pattern also led to significant artefacts in the produced images that mimic the structure of the pattern. This has motivated us to investigate alternate avenues in terms of CS-reconstruction of QAM images.


\section{AMP-Net for QAM reconstruction}
With the proliferation of deep learning and convolutional neural network techniques, interest has grown in novel neural network architectures that offer less of a black-box approach and whose inner workings can be more formally analysed. One major technique to come out of this has been deep unfolding \cite{Gregor2010} \cite{Hershey2014} \cite{Monga2021}, or deep unrolling, which allows for the design of a neural network whose layers effectively mimic the algorithmic steps of a sequential or iterative algorithm such as optimisation solvers. This allows for the design of interpretable, high-speed neural networks that can replace classical iterative algorithms (such as AMP \cite{Borgerding2017}) that are often computationally prohibitively expensive. 

To that end AMP-Net, a deep unfolded model of AMP, was recently proposed in the literature \cite{Zhang2021}. AMP-Net is based on unfolding of the iterative denoising process that forms the heart of AMP, and it endeavours to combine CNNs and the design of the sampling matrix to better fit the noise term in the AMP problem formulation. The floating-point sampling matrix $A$ itself is trainable, and is trained jointly with the other parameters of the designed deep unfolding model. As the processing is done in a block by block basis, a, optional trainable deblocking module $B_k(\cdot)$ is also introduced to alleviate blocking artefacts - a block diagram of its structure can be seen in Figure \ref{fig:ampnetstruct}.

In AMP-Net, the reconstruction/denoising task can be formulated as 

\begin{equation}
\tilde{x_i^k} = A^Tz_i^{k-1} + x_i^{k-1} - (A^TA-I)(\tilde{x_i} - x_i^{k-1})
\label{ampnet1}
\end{equation}

where $\tilde{x_i}$ is the vectorised form of an image block $X_i$, as produced for processing in AMP-Net by the image blocking function $\mathfrak {S}(\cdot)$. Obtaining the quantity $\tilde{x_i} - x_i^{k-1}$ allows us to achieve reconstruction via linear operations. Replacing this term $\tilde{x_i} - x_i^{k-1}$ with a non-linear trainable function $\mathfrak {N}_k(\cdot)$ we can rewrite (\ref{ampnet1}) as

\begin{equation}
x_i^k = A^Tz_i^{k-1} + x_i^{k-1} - (A^TA-I)vec(\mathfrak {N}_k(X_i^{k-1}))
\label{ampnet2}
\end{equation}

where the term $(A^TA-I)vec(\mathfrak {N}_k(X_i^{k-1}))$ can be regarded as the noise term. 

Sampling matrix optimisation is of particular interest to QAM CS reconstruction, given the difficulties in designing a suitable and practicable matrix from scratch mentioned previously. This has motivated us to investigate the use of AMP-Net, instead of classical AMP methods combined with handcrafted sampling matrices, for the reconstruction of QAM parametric maps.

\section{Experimental Results}

\begin{figure*}[t!]
\centering
\includegraphics[width=1\textwidth]{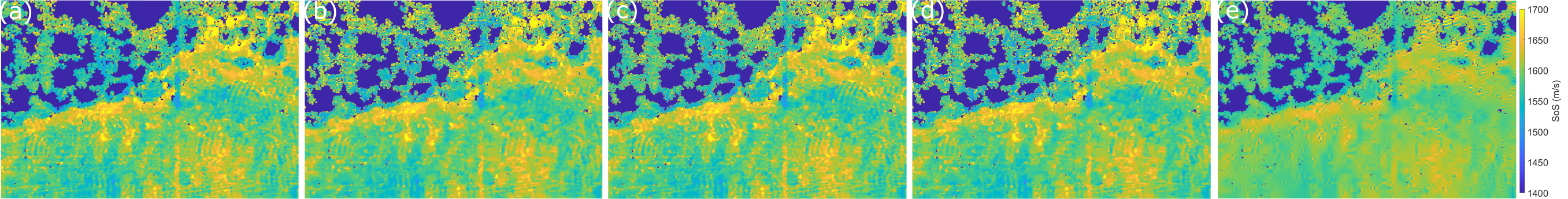}
\caption{Reconstruction of Speed of Sound (m/s) QAM maps from human lymph node data acquired at 250 MHz. (a) Ground Truth data using AR imaging \cite{Rohrbach2018}, followed by CS reconstructions using (b) AMP-Net, (c) AMP-Net-BM, (d) Q-AMP-Net and (e) Cauchy AMP \cite{Kim2018}, and 25\%  of the total samples.}
\label{fig:results1}
\end{figure*}

\begin{figure*}[t!]
\centering
\includegraphics[width=1\textwidth]{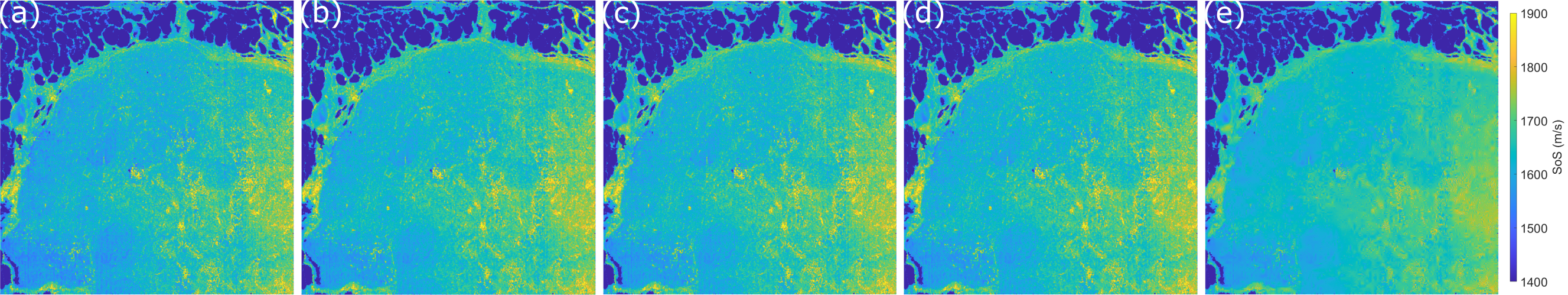}
\caption{Reconstruction of Speed of Sound (m/s) QAM maps from human lymph node data acquired at 500 MHz. (a) Ground Truth data using AR imaging \cite{Rohrbach2018}, followed by CS reconstructions using (b) AMP-Net, (c) AMP-Net-BM, (d) Q-AMP-Net and (e) Cauchy AMP \cite{Kim2018}, and 25\%  of the total samples.}
\label{fig:results2}
\end{figure*}

To evaluate AMP-Net for QAM reconstruction we perform two experiments using 250 MHz and 500 MHz QAM data. For the first experiment, we utilise a dataset of QAM data acquired using a 250-MHz system on a human lymph node thin section obtained from a colorectal cancer patient, scanned at a 2 $\mu m$ step size.This lymph node dataset contains a total of 39 images of varying sizes and aspect ratios,  and part of it was also used by the authors in \cite{Kim2018}.

The QAM dataset consists of two-dimensioal (2D) parametric maps describing various mechanical and acoustical properties of the tissue; these exhibit similar contrast and structural characteristics, and for the experiments here we used the speed of sound (SoS) maps.

We utilised three variants of AMP-Net in our experiments. The first one, denoted simply as AMP-Net, was the stock pre-trained configuration available, trained on BSDS500 as outlined in \cite{Zhang2021}. We use the AMP-Net model trained on a 25\% compression ratio and at 6 iterations (where iterations here refers to the number \textit{K} of ``unfolded'' network iterations, with options available from 2 to 9). Training is done with a batch size of 32, a learning rate of 0.0001 and over 100 epochs. The sampling matrix is initialised as a random Gaussian matrix and is not further optimised/trained in this version of AMP-Net. The deblocking module is also not utilised. This most basic version of AMP-Net can serve as a benchmark of potential performance. We then also tested using the AMP-Net-BM variant, in which the sampling matrix is jointly trainable with the reconstruction, and the optional deblocking module is engaged. For AMP-Net-BM we kept the same compression ratio (25\%) and number of iterations (6).

While the QAM lymph node dataset is small, we trained AMP-Net on a subset of it (denoted here as Q-AMP-Net). We utilised the simplest form of AMP-Net for this experiment, without a trainable sampling matrix or the deblocking module, due to the small training dataset available. We reserved 20 images for training and 9 for validation, while the remaining 10 were used for testing. Images were cropped to 321x481 pixels, the standard image size in BSDS500. Training regime was as above, again utilising the 25\% compression ratio and 6 iteration model. 

We compare the reconstructions achieved via AMP-Net, AMP-Net-BM and Q-AMP-Net to results obtained using the state of the art Cauchy AMP QAM reconstruction algorithm \cite{Kim2018}. For the Cauchy AMP experiment we have utilised a spiral sampling pattern whose spatial coverage is approximately equal to the 25\% compression ratio utilised in AMP-Net; note that due to the nature of the sampling pattern, the match in compression ratio may not be exact. The ground truth original data have been formed with a fully sampled dataset using the standard AR QAM image formation algorithm \cite{Rohrbach2018}.

Figure \ref{fig:results1} shows results obtained via the 3 AMP-Net variants operating on a sample SoS image from the QAM lymph-node dataset. The basic AMP-Net variant produces an image that is visually similar to the ground truth SoS image, with some minor loss of contrast as well as some blocking artefacts. These tend to appear primarily around edge regions of the sample. The AMP-Net-BM variant exhibits slightly better contrast, quite possibly due to the benefit of a trainable sampling matrix, while it also shows less severe blocking artefacts due to the deblocking module. Q-AMP-Net produces results that are visually near indistinguishable from the stock, BSDS500-trained version of AMP-Net. 

\begin{table}[tbp]
\caption{Numerical Evaluation of CS QAM SoS Map Reconstruction}
\begin{center}
\begin{tabular}{|c||c c c || c c c|}
\hline
  & \multicolumn{3}{|c||}{\textbf{250 MHz}} &   \multicolumn{3}{|c|}{\textbf{500 MHz}} \\
\hline
\hline
\textbf{Method} & \textbf{PSNR} & \textbf{RMSE} & \textbf{SSIM} & \textbf{PSNR} & \textbf{RMSE} & \textbf{SSIM}\\
\hline
Cauchy AMP & 17.37 & 41.3 & 0.48 & 24.47 & 14.8 & 0.19\\
\hline
AMP-Net & 25.50 & 12.6 & 0.62 & 30.16 & 7.7 & 0.58 \\
\hline
AMP-Net-BM &  \textbf{28.42} & \textbf{9.1} & \textbf{0.75}  & \textbf{32.28}  & \textbf{6} & \textbf{0.78} \\
\hline
Q-AMP-Net & 25.49 & 12.8 & 0.61 & 30.06 & 7.8 & 0.57 \\
\hline
\end{tabular}
\label{tab:results}
\end{center}
\end{table}
The three AMP-Net variants were also tested on a 500 MHz QAM acquisition of 6$\mu m$-thick human lymph node data at a 1 $\mu m$ scanning step size. This image is 1800x1800 pixels, and was tiled accordingly for processing through AMP-Net. The results are shown in Figure \ref{fig:results2} with similar trends in terms of performance to those of Figure \ref{fig:results1}. The Cauchy AMP reconstruction here appears noticeably poorer than the AMP-Net reconstructions, and the artefacts due to the spiral sampling pattern are more prominent, manifesting as concentric rings interrupting areas of high intensity. 

A quantitative assessment of the AMP-Net reconstructed images in terms of Peak Signal-to-Noise Ratio (PSNR), Root Mean Square Error (RMSE), and Structural Similarity Index (SSIM) can be found in Table \ref{tab:results}. Values quoted for the 250 MHz data are mean values across the 10 images of the dataset. AMP-Net and Q-AMP-Net score  similarly across the board, though notably the BSDS500-trained variant is consistently higher even if marginally so, showing up to 63$\%$ improvement in performance in terms of PSNR. This is perhaps not surprising given the small set of data available for the training of Q-AMP-Net, and if anything the comparable performance of the two is indicative of the overall robustness of AMP-Net. The addition of the trainable sampling matrix and the deblocking module makes AMP-Net-BM the highest performer with this being clearly indicated across all three metrics, whereas the sampling pattern-related artefacts in the Cauchy AMP reconstructions translate in poor SSIM scores.

While encouraging, this work poses a number of interesting questions for future research. AMP-Net's trainable floating point measurement matrix provides great reconstruction performance, but is not directly realisable in practice as the QAM acquisition mechanism allows only for binary spatial variation. The ability to train a binary measurement matrix, either directly or by thresholding of a floating point matrix may provide a solution to this issue. Conversely, the structured binary sampling patterns of Cauchy AMP, such as the spiral used here, can not be directly transplanted to a deeply unfolded architecture as the need for block-by-block processing with CNNs would lead to inconsistent sampling matrices.


\section{Conclusions}
We present preliminary results demonstrating the promising performance in CS QAM reconstruction using AMP-Net. Initial tests show that AMP-Net is capable of reconstructing images with better performance in terms of sharpness and contrast compared to existing CS methods, and with no adverse effects due to the sampling pattern structure. The ability to train the measurement matrix jointly with the AMP unfolded module also seems to provide a noticeable performance improvement. In the future, we aim to investigate ways of making AMP-Net more practicable by adapting to a binary measurement. We also intend to conduct more extensive experiments with a more extensive dataset of QAM imagery that will allow for more robust training, as well as more extensive ablation studies to better quantify the effects of the trainable measurement matrix and deblocking modules.

\clearpage
\bibliographystyle{IEEEbib}
\bibliography{QAM_Literature}

\end{document}